\documentclass[10pt]{iopart} 
\usepackage{amsmath,amsfonts,amssymb,amsthm,mathrsfs,csquotes}
\usepackage{graphicx,color,array,caption,booktabs}
\newenvironment{lyxcode}
{\par\begin{list}{}{\setlength{\rightmargin}{\leftmargin}
\setlength{\listparindent}{0pt}\raggedright\setlength{\itemsep}{0pt}
\setlength{\parsep}{0pt}\normalfont\ttfamily}\item[]}{\end{list}}

\makeatother

\makeatletter

\newcommand\opteq[1]{\mathrel{\mathpalette\opt@eq{#1}}}
\newcommand{\opt@eq}[2]{%
  \begingroup
  \sbox\z@{$#1#2$}%
  \sbox\tw@{\resizebox{!}{.5\ht\z@}{$\m@th#1($}}%
  \nonscript\hskip-\wd\tw@
  \mkern1mu
  \raisebox{-.35\ht\z@}[0pt][0pt]{\resizebox{!}{.5\ht\z@}{$\m@th#1($}}%
  \mkern-1mu
  {#2}%
  \mkern-1mu
  \raisebox{-.35\ht\z@}[0pt][0pt]{\resizebox{!}{.5\ht\z@}{$\m@th#1)$}}%
  \mkern1mu
  \nonscript\hskip-\wd\tw@
  \endgroup
}
\makeatother
\newcommand{\leoq}{\opteq{\leq}}

\makeatletter
\newcommand\PMopteq[1]{\mathrel{\mathpalette\PMopt@eq{#1}}}
\newcommand{\PMopt@eq}[2]{%
  \begingroup
  \sbox\z@{$#1#2$}%
  \sbox\tw@{\resizebox{!}{.5\ht\z@}{$\m@th#1($}}%
  \nonscript\hskip-\wd\tw@
  \mkern1mu
  \raisebox{-.15\ht\z@}[0pt][0pt]{\resizebox{!}{.5\ht\z@}{$\m@th#1($}}%
  \mkern-1mu
  {#2}%
  \mkern-1mu
  \raisebox{-.15\ht\z@}[0pt][0pt]{\resizebox{!}{.5\ht\z@}{$\m@th#1)$}}%
  \mkern1mu
  \nonscript\hskip-\wd\tw@
  \endgroup
}
\makeatother
\newcommand{\pmo}{\PMopteq{\pm}}

\begin{document}

\title{Back and forth from Fock space to Hilbert space: a guide for commuters}

\author{Andrea Beggi$^{1}$, Ilaria Siloi$^{2}$, Claudia Benedetti$^{3}$,
Enrico Piccinini$^{4}$, Luca Razzoli$^{5}$, Paolo Bordone$^{5}$,
Matteo G. A. Paris$^{3,6}$}

\address{$^{1}$I.I.S. {\em Primo Levi}, I-41058 Vignola, Italy.}

\address{$^{2}$Department of Physics, University of North Texas, 76201 Denton,
Texas, USA.}

\address{$^{3}$Quantum Technology Lab, Dipartimento di Fisica {\em Aldo
Pontremoli}, Universit\`{a} degli Studi di Milano, I-20133 Milano, Italy.}

\address{$^{4}$Dipartimento di Ingegneria dell'Energia Elettrica e dell'Informazione
{\em Guglielmo Marconi}, Universit\`{a} di Bologna, I-40136 Bologna,
Italy.}

\address{$^{5}$Dipartimento di Scienze Fisiche, Informatiche e Matematiche,
Universit\`{a} di Modena e Reggio Emilia, I-41125 Modena, Italy.}

\address{$^{6}$The Institute of Mathematical Sciences, CIT Campus, Taramani,
Chennai, 600113, India.}
\begin{abstract}
Quantum states of systems made of many identical particles, e.g. those
described by Fermi-Hubbard and Bose-Hubbard models, are conveniently
depicted in the Fock space. However, in order to evaluate some specific
observables or to study the system dynamics, it is often more effective
to employ the Hilbert space description. Moving effectively from one
description to the other is thus a desirable feature, especially when
a numerical approach is needed. Here we recall the construction of
the Fock space for systems of indistinguishable particles, and then
present a set of recipes and advices for those students and researchers
in the need to commute back and forth from one description to the other.
The two-particle case is discussed in some details and few guidelines
for numerical implementations are given. 
\end{abstract}
\maketitle

\section{Introduction}

The wave function describing the quantum state of a collection of
identical bosons is symmetric under the exchange of any two particles,
and is thus naturally described in the Hilbert space of symmetric
functions, which is a subspace of the tensor product of single-particle
states. On the contrary, the wave function for a collection of identical
fermions is antisymmetric, which means that it must change sign
when we exchange any two particles. These wave functions are elements
of a Hilbert space of antisymmetric functions, which is another
subspace of the tensor product of single-particle states. Indistinguishability thus introduces correlations in the wave 
function, and this is true even for non-interacting particles, a 
feature that prompted attempts to do quantum information
processing exploiting only the statistical properties of quantum 
systems \cite{oma02,bos03}. 
\par
The above facts are usually summarised by saying that for quantum 
particles with a definite statistics not all available states are 
permitted, also in those situations where one addresses free 
particles. In graduate physics courses, second quantization and
the Fock space  \cite{fck32,fw71,rs75,mah81,no88}, 
are presented as the natural framework where this constraint may be naturally taken into account. Indeed, the Fock space
is a crucial tool in the description of systems made of a variable,
or unknown, number of identical particles. 
In particular, the Fock space allows one to build the space of states starting from the single-particle Hilbert space. As a side effect,
the usual introduction of the Fock space may somehow give
the impression that the Hilbert space description may be left 
behind. On the other hand, the representation of operators
in the Fock space is not straightforward since the indexing of the
basis set states, as well as the interpretation of the number states
in terms of particle states, are usually not trivial. A known example
is that of fermionic operators on a lattice system \cite{hub63}:
anti-commutation rules have to be taken into account and additional
phases appear in the components of hopping operators from a site to
another through periodic boundary conditions. Additionally, one
often encounters operators that are symmetrized or anti-symmetrized
versions of a distinguishable particle operator, e.g. the kinetic
term in Hubbard models \cite{hub63,kol63,fis89}, and in this case
one may assume that those operators contain both bosonic and fermionic
features, which should be then discriminated (separated) using a suitable
transformation \cite{aie13}. 

For all the above reasons, it is often more transparent to employ
the Hilbert space description and to study {\em there} the dynamics
of a physical system \cite{APE}, as done in some recent works concerning
the study of quantum walks of identical particles \cite{benedetti12,beg16,siloi17,beg18}.
On the other hand, Fock number states
appear quite naturally in the description of systems of identical
particles and thus a question arises on how and whether we may go
from Fock space to Hilbert space and vice versa with minimum effort. 

The main goal of this paper is to provide a gentle introduction to
details of the transformation rules between the different description
of states and operators in the two spaces. We start smoothly, by recalling
the construction of the Fock space for systems of indistinguishable
particles, and then offer a set of recipes, guidelines, and advices
for those people interested in going back and forth from one description
to the other. We devote some attention to the two-particle case, which
already contains most of the interesting features related to indistinguishability,
and briefly discuss how to take care of the two different representations
in numerical implementations. The material presented in this paper
is intended to be a concise reference about the different representations
employed in many-body physics, and it aims at being useful to students
and researchers working with systems of identical particles, ranging
from photons in a black-body cavity to interacting electrons in a
lattice, and from neutrons in a neutron star to helium atoms in a
superfluid. 

The paper is structured as follows. In Section \ref{s:ipfs} we recall
few basic notions about indistinguishable particles and the construction
of the Fock space. In Section \ref{s:fhtf} we illustrate in details
how to change description from Hilbert space to Fock space and vice versa
in the operator representation and the system evolution. Section \ref{s:app}
presents some specific applications, whereas Section \ref{s:num}
contains guidelines to numerical implementations. Finally, Section
\ref{s:out} closes the paper with some concluding remarks. 

\section{Identical particles and the Fock space}

\subsection{From distinguishable to indistinguishable particles}

\label{s:ipfs} Let us start by considering a collection of $N$ identical
but distinguishable particles, each of which can be put in one of
the $K$ {\em modes} of a quantum system, e.g. the $K$ eigenstates
of its Hamiltonian. The collective state describing the system is
given by 
\begin{align}
\left|\Psi\right\rangle =\left|k_{1}\right\rangle \otimes\left|k_{2}\right\rangle \otimes...\otimes\left|k_{N}\right\rangle =\left|k_{1},k_{2},...,k_{N}\right\rangle ,\label{eq:Stati_Hilbert}
\end{align}
or by any linear combination of states of this kind, which all belong
to the $K^{N}$-dimensional $N$-particles Hilbert space $\mathscr{H}_{N}=\mathscr{H}_{1}^{\otimes N}$
given by the tensor product of $N$ single-particle spaces $\mathcal{\mathscr{H}}_{1}$,
each one with dimension $K$ and basis $\{\left|k_{i}\right\rangle \}_{i}$. 

Let us now introduce the notion of indistinguishability \cite{sub14}.
We start by the definition of the permutation operator $\hat{P}_{ij}$,
whose effect is to exchange the states of the particles $i$ and $j$
inside any state $\left|\{k_{i}\}_{i}\right\rangle $: 
\begin{align}
\hat{P}_{ij}\left|k_{1},k_{2},...,k_{i},...,k_{j},...,k_{N}\right\rangle =\left|k_{1},k_{2},...,k_{j},...,k_{i},...,k_{N}\right\rangle .
\end{align}
If the particles are indistinguishable, the overall state of the system
$\left|\Psi\right\rangle _{id}$ will be given by linear combination
of states (of distinguishable particles) which is invariant under
action of $\hat{P}_{ij}$, e.g. states like $\left|\{k_{i}\}_{i}\right\rangle $.
It means that $\hat{P}_{ij}\left|\Psi\right\rangle _{id}$ is a state
physically indistinguishable from the previous one, i.e. they can
differ only for a phase: 
\begin{align}
\hat{P}_{ij}\left|\Psi\right\rangle _{id}=e^{\mathrm{i}\phi}\left|\Psi\right\rangle _{id}.
\end{align}
Of course, two identical permutations must reproduce the initial state,
i.e.: 
\begin{align}
\hat{P}_{ij}^{2}\left|\Psi\right\rangle _{id}=e^{2\mathrm{i}\phi}\left|\Psi\right\rangle _{id}=\left|\Psi\right\rangle _{id},
\end{align}
and thus the eigenvalues for $\hat{P}_{ij}$ are given by $e^{\mathrm{i}\phi}=\pm1$. 

According to the spin-statistics theorem we have two categories of
identical particles: \emph{fermions}, which are characterized by half-integer
spins and anti-symmetric wavefunctions, and \emph{bosons}, which
are characterized by integer spins and symmetric wavefunctions.
High precision experiments have confirmed the spin-statistics and
established strict probability bounds for a violation to occur \cite{ss1,ss2,ss3,ss4}.
Alternative para-statistics have been suggested earlier in the history
of quantum mechanics \cite{gre53}, however we are not discussing
here the properties of those kind of particles, e.g. anyons \cite{wil82}. 

A state is symmetric or anti-symmetric under the action of $\hat{P}_{ij}$
if, respectively, it maintains or it changes its sign, i.e.: 
\begin{align}
\hat{P}_{ij}\left|\Psi\right\rangle _{F} & =-\left|\Psi\right\rangle _{F},\label{eq:Perm_FERM}\\
\hat{P}_{ij}\left|\Psi\right\rangle _{B} & =+\left|\Psi\right\rangle _{B}.\label{eq:Perm_BOSON}
\end{align}
A symmetric or anti-symmetric state can be built with the proper
symmetrization operator $\hat{S}$ or $\hat{A}$, acting on the distinguishable
particle state $\left|\Psi\right\rangle $: 
\begin{align}
\hat{S}\left|\Psi\right\rangle  & =\left|\Psi\right\rangle _{B},\\
\hat{A}\left|\Psi\right\rangle  & =\left|\Psi\right\rangle _{F}.
\end{align}
In general, the symmetrization operators can be built as 
\begin{align}
\left[\begin{array}{c}
\hat{S}\\
\hat{A}
\end{array}\right]\left|\Psi\right\rangle =\left[\begin{array}{c}
\left|\Psi\right\rangle _{B}\\
\left|\Psi\right\rangle _{F}
\end{array}\right]=\sqrt{\frac{n_{1}!n_{2}!...n_{K}!}{N!}}\sum_{\hat{P}}(\pm1)^{\sigma(P)}\hat{P}\left|\Psi\right\rangle ,
\end{align}
where we are applying on $\left|\Psi\right\rangle $ all the possible
distinct permutations $\hat{P}$ of the $N$ single-particle states
$\left|k_{i}\right\rangle $ included in $\left|\Psi\right\rangle $,
each one multiplied by the sign of the permutation $(-1)^{\sigma(P)}$,
where $\sigma(P)$ is the number of particle exchanges occurred in the permutation $\hat{P}$,
when we are dealing with fermions. Notice that any $N$-particle permutation
$\hat{P}$ can be built by composing a proper sequence of two-particle
permutations $\hat{P}_{ij}$, and the same is true for the operators
$\hat{A}$ and $\hat{S}$. The states should be then properly normalized
with the prefactor under square root, where $n_{k}$ indicates the
number of particles occupying the state $k$, i.e., the number of
times that $k$ occurs in $\left|\Psi\right\rangle $. 

>From the anti-symmetrization procedure for fermions, we derive the
\emph{Pauli exclusion principle}, that forbids two fermions to occupy
the same state: indeed, if this was the case, e.g. for the particles
$i$ and $j$, we would have contemporarily 
\begin{align}
\hat{P}_{ij}\left|\Psi\right\rangle _{F} & =-\left|\Psi\right\rangle _{F},\\
\hat{P}_{ij}\left|\Psi\right\rangle _{F} & =\left|\Psi\right\rangle _{F},
\end{align}
where the first equality comes from Eq. (\ref{eq:Perm_FERM}), while
the second one comes from the fact that the two particles occupy the
same state. The only possible conclusion is that our state is the
null vector. 

\begin{figure}[h!]
\begin{raggedleft}
\includegraphics[width=0.85\columnwidth]{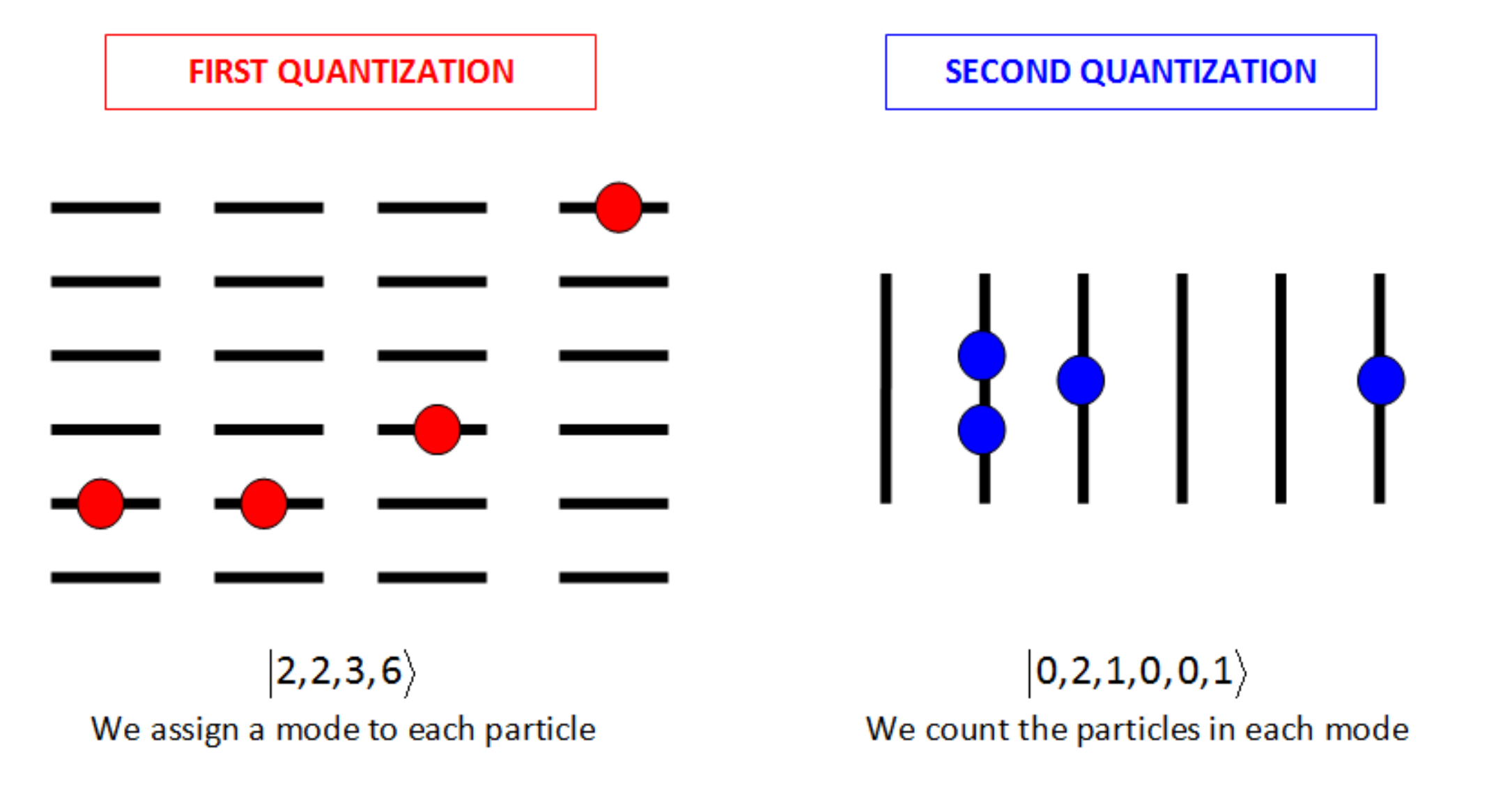} 
\par\end{raggedleft}

\caption{First and Second Quantization compared for $N=4$ particles and $K=6$
modes available to each particle.}
\label{f:1q2q} 
\end{figure}

In order to properly describe states and operators taking into account
the indistinguishability of particles we should move to the formalism
of second quantization \cite{shch2013}, where states belong to the bosonic or fermionic
Fock space $\mathscr{F}$, that is a space containing states with
a number of particles that in principle is not fixed. The \emph{number
states} (basis states) of the Fock space can be represented as 
\begin{align}
\left|n_{1},n_{2},...,n_{K}\right\rangle _{B(F)},\label{eq:Stati_Fock}
\end{align}
with the fundamental constraint $n_{i}\in\{0,1\}$ holding only for
fermions, because of Pauli's principle. If we deal with a fixed number
$N$ of particles, there is the additional constraint $\sum_{i=1}^{K}n_{i}=N$.
In this last case, we are operating in the subspace of $\mathscr{F}$
called $\mathscr{F}_{N}$. Indeed the Fock space is given by: 
\begin{align}
\mathscr{F}=\bigoplus_{N=0}^{\infty}\mathscr{F}_{N}.
\end{align}
These number states coincide with the states of Eq. (\ref{eq:Stati_Hilbert})
except for the relabeling (and the symmetrization). Indeed, while
in the \emph{first quantization} formalism we specify for each particle
$i$ ($i=1,...,N$) the state/mode $k_{i}$ that it occupies, in the
second quantization formalism we treat particles as excitation of
the modes of a field, therefore for each mode $i$ ($i=1,...,K$)
we specify how many excitation/particles $n_{i}$ it contains, since
we cannot distinguish among them (see Fig. \ref{f:1q2q}). 


The total number of particles $N$ can be changed through the application
of \emph{creation} and \emph{annihilation} (or \emph{destruction})
operators, respectively denoted as $\hat{a}_{i}^{\dagger}$ and
$\hat{a}_{i}$. These operators create or destroy a particle in the
mode/state $i$, so they connect the Fock subspaces with a different
number of particles: 
\begin{align}
\hat{a}_{i}^{\dagger} & :\mathscr{F}_{N}\rightarrow\mathscr{F}_{N+1},\\
\hat{a}_{i} & :\mathscr{F}_{N}\rightarrow\mathscr{F}_{N-1}.
\end{align}
In more detail, these operators are defined by the following commutation
rules: 
\begin{align}
\left[\hat{a}_{i},\hat{a}_{j}\right]_{\mp} & =\left[\hat{a}_{i}^{\dagger},\hat{a}_{j}^{\dagger}\right]_{\mp}=0,\\
\left[\hat{a}_{i},\hat{a}_{j}^{\dagger}\right]_{\mp} & =\delta_{i,j},
\end{align}
where the upper sign, denoting the commutator, holds for bosons,
while the lower sign, denoting the anti-commutator, holds for fermions.
It is quite important to stress, therefore, that fermionic creation
(annihilation) operators do anticommute: therefore, when we exchange
their order, we have to add a minus sign for each permutation we perform. 

Upon denoting by $\left|0\right\rangle $ the \emph{vacuum state},
corresponding to a state with no particle, i.e. $\left|0,0,...,0\right\rangle $
(not to be confused with the null vector), we can build any number
state as 
\begin{align}
\left|n_{1},n_{2},...,n_{K}\right\rangle _{B(F)}=\prod_{i=1}^{K}\frac{1}{\sqrt{n_{i}!}}\left(\hat{a}_{i}^{\dagger}\right)^{n_{i}}\left|0\right\rangle .
\end{align}
From the commutation rules, we also deduce the action of the creation
and annihilation operators on the number states: 
\begin{align}
\hat{a}_{i}^{\dagger}\left|n_{1},n_{2},...,n_{i},...,n_{K}\right\rangle _{B(F)} & =\begin{cases}
\sqrt{n_{i}+1}\left|n_{1},n_{2},...,n_{i}+1,...,n_{K}\right\rangle _{B}\\
(1-n_{i})(-1)^{\sigma_{i}}\left|n_{1},n_{2},...,1-n_{i},...,n_{K}\right\rangle _{F} \, ,
\end{cases}\label{eq:a_dagger}\\
\hat{a}_{i}\left|n_{1},n_{2},...,n_{i},...,n_{K}\right\rangle _{B(F)} & =\begin{cases}
\sqrt{n_{i}}\left|n_{1},n_{2},...,n_{i}-1,...,n_{K}\right\rangle _{B}\\
n_{i}(-1)^{\sigma_{i}}\left|n_{1},n_{2},...,1-n_{i},...,n_{K}\right\rangle _{F} \, ,
\end{cases}\label{eq:a}
\end{align}
where the $\sigma_{i}$ exponent is due to the anti-commuting rules and
is given by 
\begin{align}
\sigma_{i}=\sum_{k=1}^{i-1}n_{k}.
\end{align}
From Eqs. (\ref{eq:a_dagger})-(\ref{eq:a}) we straightforwardly
deduce that 
\begin{align}
\hat{a}_{i}\left|n_{1},n_{2},...,n_{i},...,n_{K}\right\rangle  & =0\quad\mathrm{if}\, n_{i}=0,\\
\hat{a}_{i}^{\dagger}\hat{a}_{i}\left|n_{1},n_{2},...,n_{i},...,n_{K}\right\rangle  & =n_{i}\left|n_{1},n_{2},...,n_{i},...,n_{K}\right\rangle .
\end{align}
The last equation gives the definition of the \emph{number operator}
\begin{align}
\hat{n}_{i}=\hat{a}_{i}^{\dagger}\hat{a}_{i},
\end{align}
that counts the number of particles in the mode $i$. 

Let us now consider a single-particle operator $\hat{O}_{j}$ acting
on the particle $j$. It is apparent that we must have 
\begin{align}
\hat{P}_{ij}\hat{O}_{j}\hat{P}_{ij}=\hat{O_{i}}.
\end{align}
Then, for a many-particle operator $\hat{O}$, we will say that this
operator is invariant under particle exchange (\emph{permutation symmetry})
if, for any couple of particles $i$ and $j$, the following relation
holds: 
\begin{align}
\hat{P}_{ij}\hat{O}\hat{P}_{ij}=\hat{O}.
\end{align}
However, the previous property does not mean that the operator $\hat{O}$
must be invariant also under the action of $\hat{A}$ and $\hat{S}$,
but it implies that these symmetry operators must commute with $\hat{O}$.
It means, therefore, that for $\hat{O}=\hat{H}$ the evolution preserves
the symmetry of bosonic and fermionic states, i.e. their subspaces
\begin{equation}
\mathscr{F}_{N}^{B}=\hat{S}\mathscr{H}_{N}
\end{equation}
and
\begin{equation}
\mathscr{F}_{N}^{F}=\hat{A}\mathscr{H}_{N}
\end{equation}
are not mixed, since the Hamiltonian $\hat{H}$ and $\hat{S}\,(\hat{A})$
have a common set of orthogonal eigenstates. Therefore, this kind
of operators can be used in the second quantization formalism with
identical particle states in the same way they were used in first
quantization, because they preserve the permutation symmetry. 
We remind here that second quantization representation is strictly
connected with the choice of the basis $\{ \left|a_{i}\right\rangle\}_{i} $
in which the operators $\hat{a}_{i}^{\dagger}\,(\hat{a}_{i})$ are
creating (destroying) particles: indeed, if $\hat{a}_{i}^{\dagger}\,(\hat{a}_{i})$
creates (destroys) a particle in the \emph{i}-th eigenstate of the
Hamiltonian, we can define a change of basis and build a new set of
operators $\hat{b}_{i}^{\dagger}\,(\hat{b}_{i})$ that create (destroy)
particles in the \emph{i}-th eigenstate of any other operator $\hat{O}$
(e.g. in the \emph{i}-th site of a lattice for the position operator): 
\begin{align}
\left|b_{i}\right\rangle =\left[\hat{b}_{i}^{\dagger}\right]\left|0\right\rangle =\sum_{j}\left\langle a_{j}\middle|b_{i}\right\rangle \left|a_{j}\right\rangle =\left[\sum_{j}\left\langle a_{j}\middle|b_{i}\right\rangle \hat{a}_{j}^{\dagger}\right]\left|0\right\rangle .
\end{align}

\subsection{Two-particle case}

Here and in the rest of the article we apply the tools introduced in the previous section
to the specific case of $N=2$ identical particles. Despite the small
number of particles, this example shows most of the peculiarities related
to indistinguishability, and illustrates how to deal with both Fock
and Hilbert descriptions to conveniently describe the physics of the
system. 

\begin{figure}[h!]
\begin{raggedleft}
\includegraphics[width=0.7\columnwidth]{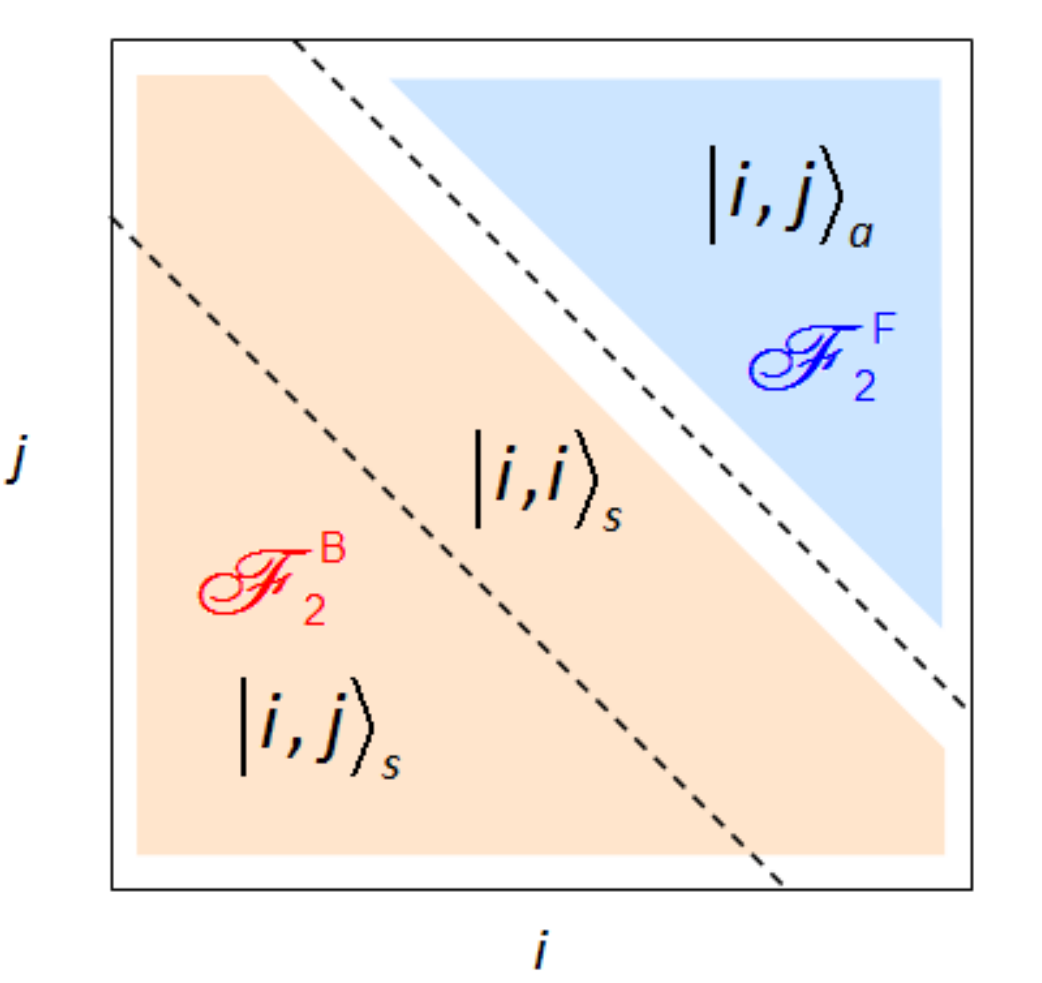} 
\par\end{raggedleft}

\caption{Hilbert space decomposition into symmetry-defined Fock subspaces for
$N=2$ particles.}
\label{f:dec} 
\end{figure}

In the Hilbert space $\mathscr{H}_{2}$ (for distinguishable particles),
the basis set is given by $\{\left|i,j\right\rangle\}_{i,j} $. According to
the symmetrization procedures described above for a 2-boson state
we have 
\begin{align}
\left|i,j\right\rangle _{s}=\begin{cases}
\frac{1}{\sqrt{2}}\left(\left|i,j\right\rangle +\left|j,i\right\rangle \right) & \mathrm{for}\, i<j\\
\left|i,i\right\rangle  & \mathrm{for}\, i=j \, ,
\end{cases}
\end{align}
while for a 2-fermion state the allowed basis set is given by: 
\begin{align}
\left|i,j\right\rangle _{a}=\frac{1}{\sqrt{2}}\left(\left|i,j\right\rangle -\left|j,i\right\rangle \right)\quad\mathrm{for}\, i<j,
\end{align}
where the constraint $j>i$ avoids overcounting in the basis set,
since 
\begin{align}
\left|j,i\right\rangle _{s} & =\left|i,j\right\rangle _{s},\\
\left|j,i\right\rangle _{a} & =-\left|i,j\right\rangle _{a}.
\end{align}
It is easy to check that these two new basis sets are orthogonal and
related to the previous one by
\begin{align}
\{\left|i,j\right\rangle \}_{i,j}=\{\left|i,j\right\rangle _{s}\}_{i,j}\cup\{\left|i,j\right\rangle _{a}\}_{i,j}.\label{eq:Basis_2_part_Symm}
\end{align}
Indeed, the dimension $d_{B}$ of the basis set for bosonic particles
is given by 
\begin{align}
d_{B}=K+\frac{K(K-1)}{2}=\frac{K(K+1)}{2},
\end{align}
while the dimension $d_{F}$ of the fermionic basis set is given by
\begin{align}
d_{F}=\frac{K(K-1)}{2},
\end{align}
and 
\begin{align}
d_{B}+d_{F}=K^{2}=\dim(\mathscr{H}_{2}).
\end{align}
Overall, the Hilbert space $\mathscr{H}_{2}$ is decomposed into two
subspaces \cite{Decomp}: one contains only symmetric states, while
the other one only anti-symmetric states, see Fig. \ref{f:dec}. These
subspaces are, in turn, the 2-particle restrictions of the Fock spaces
for bosons and fermions, namely $\mathscr{F}_{2}^{B}$ and $\mathscr{F}_{2}^{F}$:
\begin{align}
\mathscr{H}_{2}=\mathscr{F}_{2}^{B}\oplus\mathscr{F}_{2}^{F}.
\end{align}

Each subspace is obtained, in general, by applying the proper symmetry
operators over $\mathscr{H}_{2}$: 
\begin{align}
\mathscr{F}_{2}^{B} & =\hat{S}\mathscr{H}_{2},\\
\mathscr{F}_{2}^{F} & =\hat{A}\mathscr{H}_{2},
\end{align}
which are given by the following projectors: 
\begin{align}
\hat{S} & =\sum_{i}\left|i,i\right\rangle \left\langle i,i\right|+\sum_{i,j>i}\frac{1}{2}\left(\left|i,j\right\rangle +\left|j,i\right\rangle \right)\left(\left\langle i,j\right|+\left\langle j,i\right|\right),\\
\hat{A} & =\sum_{i,j>i}\frac{1}{2}\left(\left|i,j\right\rangle -\left|j,i\right\rangle \right)\left(\left\langle i,j\right|-\left\langle j,i\right|\right).
\end{align}
However, the action of $\hat{S}$ and $\hat{A}$ is not enough to
write operators (or states) in the Fock space. Indeed, if we consider
the bosonic operator $\hat{O}_{B}=\hat{S}\hat{O}\hat{S}$, in order
to properly represent it in the Fock space we still have to perform
a change of basis. Therefore, sometimes it could be useful to perform
both operations in a single step, by writing the symmetry operators
in a mixed-basis representation (where the bras are states of $\mathscr{H}_{2}$,
while kets are states of $\mathscr{F}_{2}$): 
\begin{align}
\hat{S} & =\sum_{i}\left|i,i\right\rangle _{s}\left\langle i,i\right|+\sum_{i,j>i}\left|i,j\right\rangle _{s}\frac{1}{\sqrt{2}}\left(\left\langle i,j\right|+\left\langle j,i\right|\right),\\
\hat{A} & =\sum_{i,j>i}\left|i,j\right\rangle _{a}\frac{1}{\sqrt{2}}\left(\left\langle i,j\right|-\left\langle j,i\right|\right).
\end{align}
so that we can write the operator $\hat{O}_{B}$ in $\mathscr{F}_{2}$
directly as $\hat{S}\hat{O}\hat{S}^{\dagger}$ (a similar discussion
also holds for fermionic operators).

\begin{figure}[h!]
\begin{raggedleft}
\includegraphics[width=0.8\columnwidth]{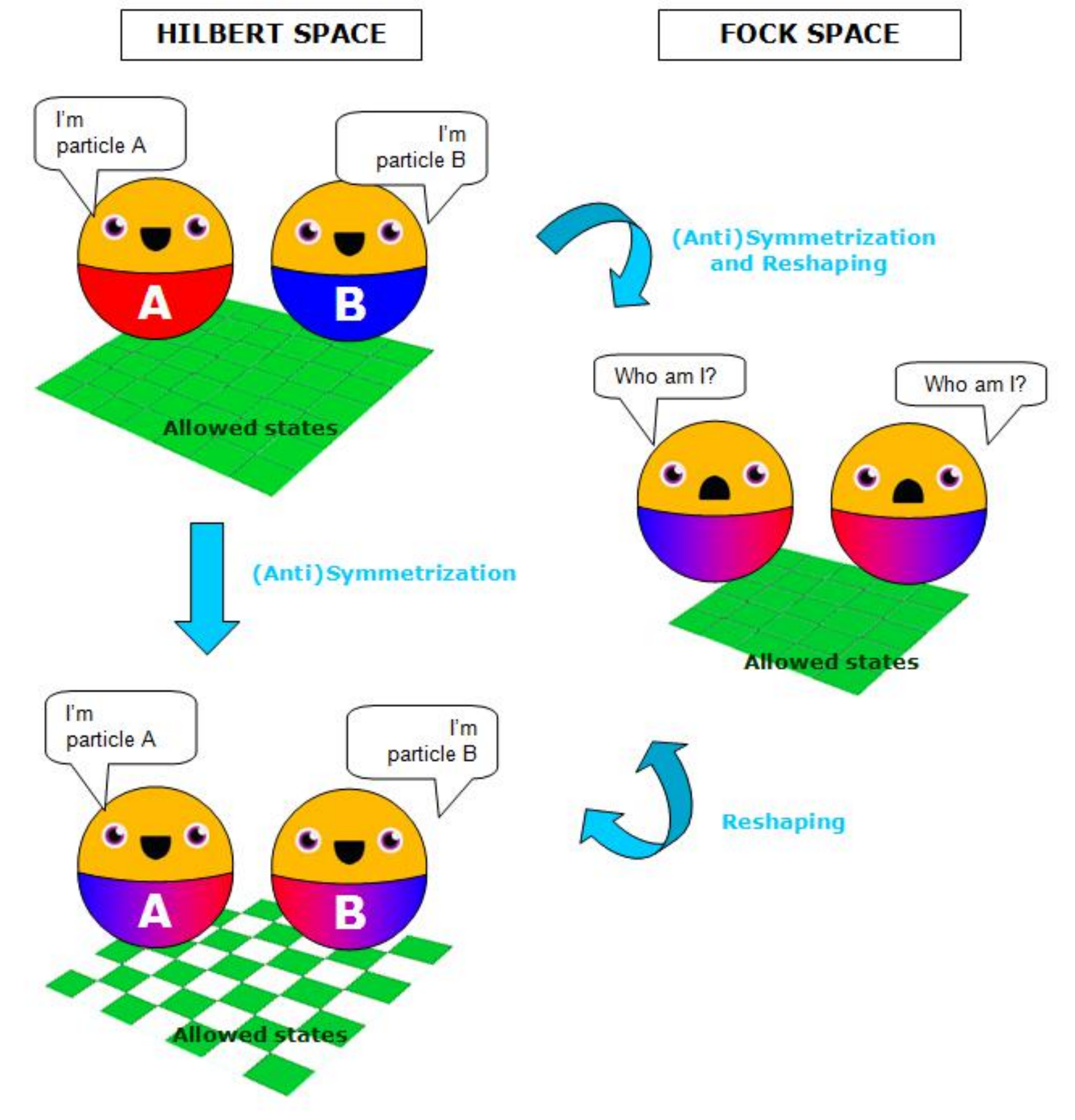} 
\par\end{raggedleft}

\caption{Pictorial representation of \emph{symmetrization} and \emph{reshaping}
processes, which transform states and operators, respectively, from
Hilbert to Fock space and vice versa. Notice that in the Hilbert space
all particles have a clear identity (A or B in the example) and may
even possess well defined states (the colors in the example) if they
are distinguishable, whereas in the Fock space no particle has a clear
identity or a well defined state. Switching from Hilbert to Fock space
requires to properly (anti-)symmetrize states and to remove states with
the wrong symmetry (reshaping), while going back to Hilbert space requires
to reshape both operators and states, but the number of allowed states
remains the same. }

\label{fig:Hamtaro-hilbert-fock} 
\end{figure}

Even if in general this is not true, we observe that for the case $N=2$ we have
\begin{align}
\hat{S}+\hat{A}=\hat{I},
\end{align}
in agreement with Eq. (\ref{eq:Basis_2_part_Symm}). In conclusion,
a distinguishable-particle operator $\hat{O}$ in $\mathscr{H}_{2}$
is invariant under particle-exchange symmetry if and only if the following
decomposition holds: 
\begin{align}
\hat{O}=\hat{S}\hat{O}\hat{S}+\hat{A}\hat{O}\hat{A}=\hat{O}_{s}+\hat{O}_{a},
\end{align}
i.e. the operator is the sum of its projections over the bosonic
and fermionic subspaces of $\mathscr{H}_{2}$. This means that $\hat{O}$
does not mix states belonging to subspaces with different symmetries
(i.e., fermionic and bosonic), and it commutes with $\hat{A}$ and
$\hat{S}$. If this property holds also for the Hamiltonian $\hat{H}$,
it is possible to perform a symmetrization only over the state vectors
without modifying the operators. Indeed, $\hat{S}$ and $\hat{A}$
are projectors, $\hat{S}=\hat{S}^{\dagger}$ and $\hat{S}^{n}=\hat{S}$,
and are orthogonal, i.e. $\hat{S}\hat{A}=\hat{A}\hat{S}=0$. The dynamics
thus conserves symmetries, and this is sufficient to get the right
expectation values: 
\begin{equation}
\hat{H}_{s} =\hat{S}\hat{H}\hat{S}=\hat{H}\hat{S}^{2}=\hat{H}\hat{S},
\end{equation}
\begin{equation}
\hat{H}_{s}\left|\Psi\right\rangle _{s} =\hat{H}\hat{S}^{2}\left|\Psi\right\rangle =\hat{H}\left|\Psi\right\rangle _{s},
\end{equation}
\begin{equation}
O_{s}={}_{s}\langle \Psi\vert\hat{O}_{s}\vert\Psi\rangle _{s} =\left\langle \Psi\right|\hat{S}^{2}\hat{O}\hat{S}^{2}\left|\Psi\right\rangle =\left\langle \Psi\right|\hat{S}\hat{O}\hat{S}\left|\Psi\right\rangle ={}_{s}\langle \Psi\vert\hat{O}\vert\Psi\rangle _{s}.
\end{equation}
Clearly for the 2-particle case, the evaluation of expectation values
and dynamics of the system can be conveniently obtained in the Hilbert
space, starting with a properly (anti-)symmetrized state\cite{Alt_dyn}, namely $\left|\Psi\right\rangle _{(a)s}$.
Then, with a proper \emph{reshaping operation}, observables may
be recast in the Fock space, see Fig. \ref{fig:Hamtaro-hilbert-fock}.
This procedure presents advantages compared to a direct calculation
in the Fock space where each particle loses its identity. For instance,
the operation of partial trace over the degrees of freedom of one
particle is straightforward in the Hilbert space, while it is more
delicate in the Fock space, where the natural basis set is the occupation
number. Moreover, the indexing of the basis in the Fock space is not
trivial to handle \cite{zhan10}, and a convenient way to rank the
basis vectors should be considered in the numerical implementation
(see Section 5). Therefore, it is often the case that the dynamics
and all the required observables are evaluated in the Hilbert space,
after a proper symmetrization of the initial state. 


\section{From Hilbert space to Fock space and vice versa}

\label{s:fhtf}

\subsection{Operator representation: from Fock to Hilbert}

Given the basis sets for the symmetric and antisymmetric subspaces,
$\{\left|i,j\right\rangle _{s}\}_{i,j\ge i}$ and $\{\left|i,j\right\rangle _{a}\}_{i,j>i}$,
their union $\mathcal{B}^{*}=\{\left|i,j\right\rangle _{s}\}_{i,j\ge i}\cup\{\left|i,j\right\rangle _{a}\}_{i,j>i}$
is a basis set for the whole Hilbert space $\mathscr{H}_{2}$. As
discussed in the previous section, we typically use the distinguishable-particle
basis set, hereafter labeled $\mathcal{B}=\{\left|i,j\right\rangle \}_{i,j}$. 

The bosonic operator $\hat{O}_{B}$ may be represented with a $d_{B}\times d_{B}$
matrix defined on $\mathscr{F}_{2}^{B}$, but also with a $K^2\times K^2$
matrix acting on $\mathscr{H}_{2}$. Adding $d_{F}$ rows and $d_{F}$
columns full of zeros, we may extend the Fock matrix $\hat{O}_{B}^{\mathscr{F}}$:
the first $d_{B}$ rows and columns involve only bosonic states, while
the additional lines only the fermionic states (none, since it is
a bosonic operator): 

\begin{align}
\hat{O}_{B}^{\mathscr{H}}|_{\mathcal{B}^{*}}=\left(\begin{array}{ccc|c}
 &  &  &  \\
 & \text{\huge{$\hat{O}_{B}^{\mathscr{F}}$}} &  & \text{\huge{$0$}} \\
 &  &  &  \\ \hline
 &  &  &  \\
 & \text{\huge{$0$}} &  & \text{\huge{$0$}} 
\end{array}\right).
\end{align}

When the basis set is given by $\mathcal{B}^{*}$, whose symmetric part coincides
with the basis of $\mathcal{\mathscr{F}}_{2}^{B}$, the representation $\hat{O}_{B}^{\mathcal{\mathscr{H}}}|_{\mathcal{B}^{*}}$
is valid in $\mathscr{H}_{2}$. In order to represent operators in
the basis $\mathcal{B}$ of distinguishable particles, we need to
reverse this transformation, that is: 
\begin{align}
\hat{O}_{B}^{\mathcal{\mathscr{H}}}|_{\mathcal{B}}=\hat{S}^{\dagger}\hat{O}_{B}^{\mathcal{\mathscr{H}}}|_{\mathcal{B}^{*}}\hat{S}.
\end{align}
The relation between the elements of a bosonic operator in the Fock
(basis $\mathcal{B}_{s}^{*}=\{\left|i,j\right\rangle _{s}\}_{i,j\ge i}$)
and in the Hilbert space (basis $\mathcal{B}=\{\left|i,j\right\rangle \}_{i,j}$)
is the following:
\begin{align}
\left(O_{B}^{\mathcal{\mathscr{H}}}\right)_{i,j;k,l} & =\left\langle i,j\right|\hat{O}_{B}^{\mathcal{\mathscr{H}}}|_{\mathcal{B}}\left|k,l\right\rangle =\frac{1}{2}\left(O_{B}^{\mathscr{F}}\right)_{i,j;k,l}\left(1+\epsilon\delta_{i,j}\right)\left(1+\epsilon\delta_{k,l}\right),\label{eq:F2Hil_bosons}
\end{align}
where $\epsilon=\sqrt{2}-1$. 

Analogous arguments apply to the fermionic case, thus 
\begin{align}
\hat{O}_{F}^{\mathscr{H}}|_{\mathcal{B}}=\hat{A}^{\dagger}\hat{O}_{F}^{\mathscr{H}}|_{\mathcal{B}^{*}}\hat{A},
\end{align}
where 

\begin{align}
\hat{O}_{F}^{\mathscr{H}}|_{\mathcal{B}^{*}}=\left(\begin{array}{c|ccc}
 \text{\huge{$0$}} &  & \text{\huge{$0$}} & \\ \hline
 \\
\text{\huge{$0$}} &  & \text{\huge{$\hat{O}_{F}^{\mathscr{F}}$}}\\
\\
\end{array}\right).
\end{align}

Again, the transformation of a fermionic operator in the Fock space
(basis $\mathcal{B}_{a}^{*}=\{\left|i,j\right\rangle _{a}\}_{i,j>i}$)
to the Hilbert space (basis $\mathcal{B}=\{\left|i,j\right\rangle \}_{i,j}$)
is given by 
\begin{align}
\left(O_{F}^{\mathscr{H}}\right)_{i,j;k,l}=\frac{1}{2}\left(O_{F}^{\mathscr{F}}\right)_{i,j;k,l}\left(1-\delta_{i,j}\right)\left(1-\delta_{k,l}\right)\varsigma_{i,j}\varsigma_{k,l},\label{eq:F2Hil_fermions}
\end{align}
where $\varsigma_{i,j}=\mathrm{sgn}(j-i)$. 

In the case of a 2-particle bosonic (fermionic) operator, the transformation
laws from Fock to Hilbert space of distinguishable particles are thus
given by equations (\ref{eq:F2Hil_bosons}) and (\ref{eq:F2Hil_fermions}),
respectively. However, one should remind that the Fock space elements
exist only for $i\le j$ ($i<j$ for fermions), so indices in $\left(O_{F}^{\mathscr{F}}\right)_{i,j;k,l}$
must be exchanged if $i>j$ and/or $k>l$. Moreover, if $i=j$ or
$k=l$, the corresponding elements are zero.

Notice that if the dynamics is evaluated in
the Hilbert space, the reshaping operation needed to recast observables
in the Fock space is given by the inverse equations of ({\ref{eq:F2Hil_bosons}})
and ({\ref{eq:F2Hil_fermions}}).{
}

\subsection{Symmetrized and antisymmetrized operators of distinguishable particles}

\begin{figure}[h!]
\begin{raggedleft}
\includegraphics[width=0.8\columnwidth]{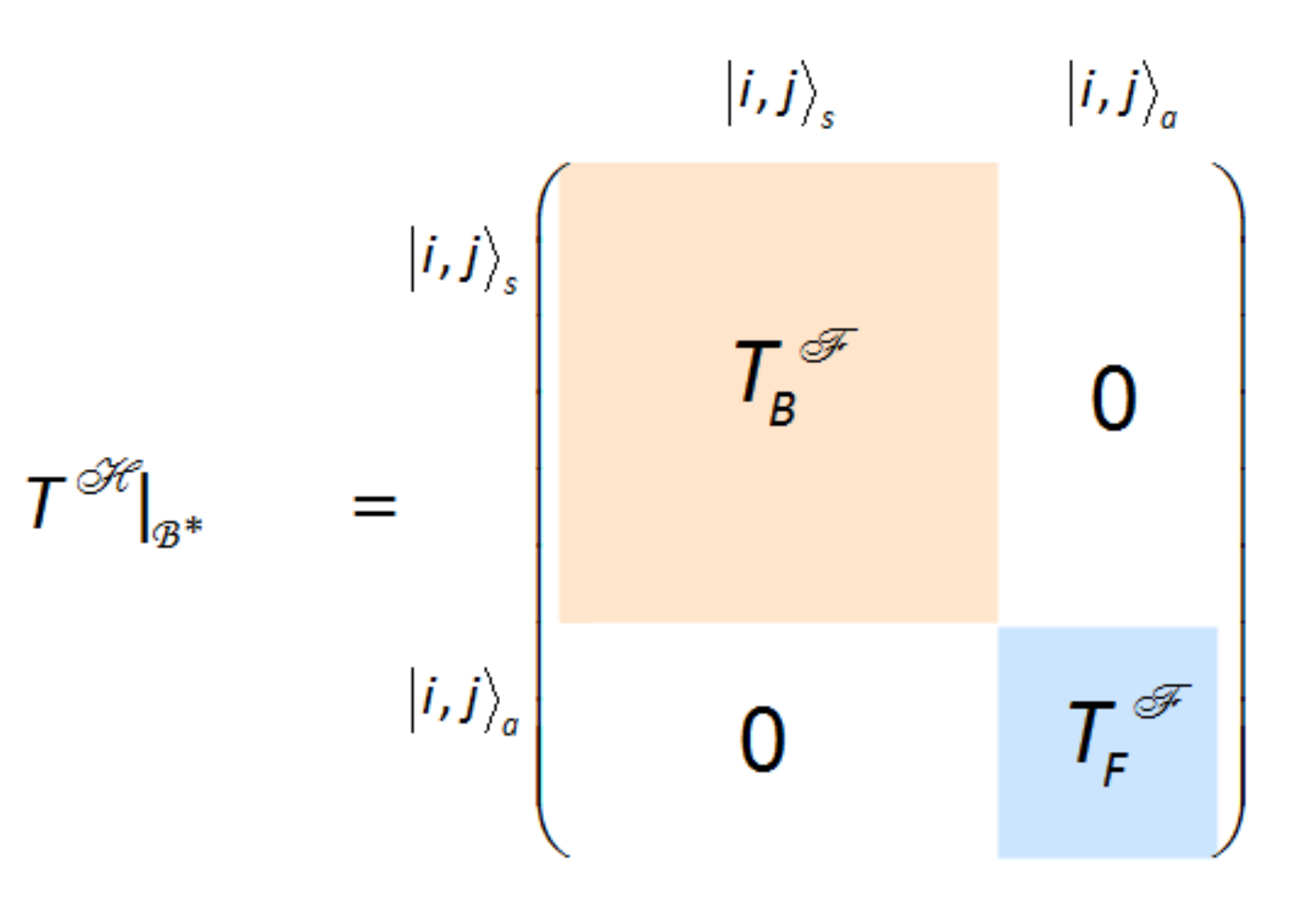} 
\par\end{raggedleft}

\caption{Representation of a parity-conserving operator: symmetry-defined subspaces
are invariant.\label{fig:Representation-of-a}}
\end{figure}

Let us consider a {\em native} Hilbert operator $\hat{T}^{\mathscr{H}}$,
i.e. an operator which arises naturally in the distinguishable particles
Hilbert space $\mathscr{H}_{2}$ where the basis set is given by $\mathcal{B}$.
Such an operator conserves parity, being invariant under particle
exchange, thus it does not mix states with different symmetries; therefore
it is equivalent to the sum of its projections with defined symmetries:
$\hat{T}^{\mathscr{H}}=\hat{S}\hat{T}^{\mathscr{H}}\hat{S}^{\dagger}+\hat{A}\hat{T}^{\mathscr{H}}\hat{A}^{\dagger}$,
see Fig. \ref{fig:Representation-of-a}. Overall, we have that $\mathscr{H}_{2}^{s}$
and $\mathscr{H}_{2}^{a}$ are invariant subspaces for $\hat{T}^{\mathscr{H}}$. 

Since the operator acts on states of distinguishable particles, it
contains both bosonic and fermionic components, which can be isolated
with a suitable transformation. An interesting example is the kinetic
term $\hat{T}$ in the Hubbard model, which can be derived from the 
discretization of the laplacian terms in the Schr\"{o}dinger equation,
and describes the hopping of
the two particles along a chain with $K$ sites. In the distinguishable
particles framework we have 
\begin{align}
\hat{T}^{\mathscr{H}}=\left[-J\sum_{i=1}^{K}(\left|i\right\rangle \left\langle i+1\right|+\left|i+1\right\rangle \left\langle i\right|)\right]\otimes I_{1}+I_{1}\otimes\left[-J\sum_{i=1}^{K}(\left|i\right\rangle \left\langle i+1\right|+\left|i+1\right\rangle \left\langle i\right|)\right],
\end{align}
where $I_{1}=\sum_{i=1}^{K}\left|i\right\rangle \left\langle i\right|$
is the single-particle identity operator, and $J$ is a scale factor that represents the tunneling amplitude between adjacent sites and depends on physical parameters of the system, such as the particle mass and the distance between the discrete sites. The form of $\hat{T}$ for
identical particles in the Fock space is 
\begin{align}
\hat{T}^{\mathscr{F}}=-J\sum_{i=1}^{K}(\hat{c}_{i}^{\dagger}\hat{c}_{i+1}+\hat{c}_{i+1}^{\dagger}\hat{c}_{i}),
\end{align}
where $\hat{c}_{i}$ is an annihilation operator (for bosons or fermions)
and $\hat{c}_{i}^{\dagger}$ the corresponding creation operator for
the mode $i$. Given the results of the previous section, we conclude
that the representation of the bosonic/fermionic operator in the Fock
space can be obtained from $\hat{T}^{\mathscr{H}}$ by a simple change
of basis, followed by a projection over the subspace with the required
symmetry. Considering $\hat{T}$ a bosonic operator, we have 
\begin{align}
\hat{T}^{\mathscr{F}}_B=\hat{S}\hat{T}^{\mathscr{H}}\hat{S}^{\dagger}.
\end{align}

Let us check the previous results in some significant cases:
\begin{align}
{}_{s}\langle k,k+1\vert \hat{T}^{\mathscr{F}}_B\vert k,k \rangle _{s} & =-J{}_{s}\langle k,k+1\vert\hat{c}_{k+1}^{\dagger}\hat{c}_{k}\vert k,k \rangle _{s}=-\sqrt{2}J,\\
{}_{s}\langle k,k+1\vert\hat{S}\hat{T}^{\mathscr{H}}\hat{S}^{\dagger}\vert k,k \rangle _{s} & =\frac{1}{\sqrt{2}}\left(\left\langle k,k+1\right|+\left\langle k+1,k\right|\right)\hat{T}^{\mathscr{H}}\left|k,k\right\rangle =-\sqrt{2}J,\\
_{s}\langle k,l+1\vert\hat{T}^{\mathscr{F}}_B\vert k,l \rangle _{s} & \underset{_{l\neq k,k-1}}{=}-J{}_{s}\langle k,l+1\vert \hat{c}_{l+1}^{\dagger}\hat{c}_{l}\vert k,l \rangle _{s}=-J,\\
_{s}\langle k,l+1\vert\hat{S}\hat{T}^{\mathscr{H}}\hat{S}^{\dagger}\vert k,l \rangle _{s} & \underset{_{l\neq k,k-1}}{=}\frac{1}{\sqrt{2}}\left(\left\langle k,l+1\right|+\left\langle l+1,k\right|\right)\hat{T}^{\mathscr{H}}\frac{1}{\sqrt{2}}\left(\left|k,l\right\rangle +\left|l,k\right\rangle \right)\notag\\
 & \quad=-J.
\end{align}
Analogous results may be obtained for fermions, where additional attention
to the anticommutation relation is required to handle periodic boundary
conditions (PBC). If a state like $\left|k,K\right\rangle _{a}$
is connected to a state like $\left|1,k\right\rangle _{a}$, e.g.
$K$ jumps over the border, we should account for an additional minus
sign due to the reordering of the anti-commuting fermionic operators.
Indeed  (we denote 
the state with no particle 
with $\left|0\right\rangle$):
\begin{align}
\left|1,k\right\rangle _{a} & =\hat{c}_{1}^{\dagger}\hat{c}_{k}^{\dagger}\left|0\right\rangle ,\\
\hat{c}_{K+1}^{\dagger}\hat{c}_{K}\left|k,K\right\rangle _{a} & \underset{_{PBC}}{=}\hat{c}_{1}^{\dagger}\hat{c}_{K}\left|k,K\right\rangle _{a}=\hat{c}_{1}^{\dagger}\hat{c}_{K}\hat{c}_{k}^{\dagger}\hat{c}_{K}^{\dagger}\left|0\right\rangle \nonumber \\
 & =-\hat{c}_{1}^{\dagger}\hat{c}_{k}^{\dagger}\hat{c}_{K}\hat{c}_{K}^{\dagger}\left|0\right\rangle =-\hat{c}_{1}^{\dagger}\hat{c}_{k}^{\dagger}\left|0\right\rangle =-\left|1,k\right\rangle _{a}.
\end{align}
The same results may be obtained without any change of basis, that
is by simply applying the operator $\hat{T}^{\mathscr{H}}$ only
over the proper (anti-)symmetrized states (this is equivalent to applying
the transformation described by $\hat{S}$ or $\hat{A}$). However,
the spectrum of the operator $\hat{S}\hat{T}^{\mathscr{H}}\hat{S}^{\dagger}$
is substantially the spectrum of the bosonic operator, while the spectrum
of $\hat{T}^{\mathscr{H}}$ contains also the fermionic eigenvalues
of $\hat{A}\hat{T}^{\mathscr{H}}\hat{A}^{\dagger}$. 

Further, we observe that the representation of $\hat{T}^{\mathscr{F}}_B$
in the basis $\mathcal{B}$ of $\mathscr{H}_{2}$ is quite interesting.
Indeed, it contains terms like $-J\vert i+1,j \rangle _{s}{}_{s}\langle i,j\vert$,
which can be rewritten as: 
\begin{align}
-J & \vert i+1,j \rangle _{s}{}_{s}\langle i,j\vert=\\
 & =-\frac{J}{2}\left(\left|i+1,j\right\rangle \left\langle i,j\right|+\left|j,i+1\right\rangle \left\langle j,i\right|+\left|j,i+1\right\rangle \left\langle i,j\right|+\left|i+1,j\right\rangle \left\langle j,i\right|\right).\notag
\end{align}
This suggests that the Fock operator not only produces transitions
where one particle hops from a site to a nearest-neighbour one, but can also
allows the two particles to exchange their position: the third term in
brackets in the RHS of the previous equation sees the second
particle in position $i$ jumping on the site $i+1$, and then exchanging
its position with the first particle, previously located on site $j$. These
\emph{exchange terms} are a consequence of the fact that when we rewrite
the operator in the Hilbert space of distinguishable particles, it
must bear signs of the exchange symmetry, due to the fact that it
is actually acting on identical particles (in this case, bosons).
The original operator $\hat{T}^{\mathscr{H}}$ does not contain these
terms, but they appear when we apply the transformation $\hat{S}\hat{T}^{\mathscr{H}}\hat{S}^{\dagger}$.

\subsection{Evolution of the system and symmetrization}

Let us now consider our quantum system of $N=2$ identical particles,
whose evolution is ruled by the Hamiltonian $\hat{H}^{\mathscr{F}}$.
Their dynamics can be directly calculated in the Hilbert space of
distinguishable particles $\mathscr{H}_{2}$: we can properly (anti-)symmetrize
the initial state and get the final state of the evolution with the
required symmetry, exactly as we had carried over the evolution in
the Fock space. This can be done either using the Fock Hamiltonian
$\hat{H}^{\mathscr{F}}$ rewritten in the Hilbert space (see Eqs.
(\ref{eq:F2Hil_bosons}) and (\ref{eq:F2Hil_fermions})), or using
directly the equivalent Hamiltonian $\hat{H}^{\mathscr{H}}$ for distinguishable
particles, provided that it conserves parity (i.e., it is invariant
under particle exchange). Indeed, projecting $\hat{H}^{\mathscr{H}}$
over the subspaces with the proper symmetry (i.e., using $\hat{S}$
or $\hat{A}$) - and/or applying it only over properly symmetrized states
- is equivalent to using $\hat{H}^{\mathscr{F}}$, as we have just
seen in the previous section. 

\section{Expectation values and projections}

\label{s:app}

\subsection{Density operator}

In order to simplify the notation, let us define a factor $g$ to
distinguish between bosons ($g=+1$) and fermions ($g=-1$).
If needed, we will use the subscript $\left|...\right\rangle _{g}$
to denote symmetrized or anti-symmetrized states. 

The density operator is the fundamental quantity for evaluating the
expectation values of all the observables characterizing the system.
The density operator can be calculated in the Hilbert space as usual:
\begin{align}
\rho^{\mathscr{H}}(t)=\left|\Psi(t)\right\rangle \left\langle \Psi(t)\right|,
\end{align}
and then recast in the Fock space using Eq. (\ref{eq:F2Hil_bosons})
for bosons ($j\ge i$, $l\ge k$): 
\begin{align}
\rho_{i,j;k,l}^{\mathscr{F}}=\frac{2}{\left(1+\epsilon\delta_{i,j}\right)\left(1+\epsilon\delta_{k,l}\right)}\rho_{i,j;k,l}^{\mathscr{H}},\label{eq:Dens_bos}
\end{align}
and Eq. (\ref{eq:F2Hil_fermions}) for fermions ($j>i$, $l>k$):
\begin{align}
\rho_{i,j;k,l}^{\mathscr{F}}=2\rho_{i,j;k,l}^{\mathscr{H}},\label{eq:Dens_fer}
\end{align}
where, within the index constraint, we have $\varsigma_{i,j}=\varsigma_{k,l}=1$.
Notice that if $\left|\Psi(t)\right\rangle =\sum_{i,j}\beta_{i,j}(t)\left|i,j\right\rangle $,
we should remember that for exchange symmetry 
\begin{align}
\beta_{i,j}(t)=g\cdot\beta_{j,i}(t),
\end{align}
and then we have 
\begin{align}
\rho^{\mathscr{H}}(t)=\sum_{i,j}\sum_{k,l}\beta_{i,j}(t)\beta_{k,l}^{*}(t)\left|i,j\right\rangle \left\langle k,l\right|.
\end{align}
If in the Fock space we have 
\begin{align}
\rho^{\mathscr{F}}(t)=\sum_{i,j\ge i}\sum_{k,l\ge k}\alpha_{i,j}(t)\alpha_{k,l}^{*}(t)\vert i,j \rangle _{g}{}_{g}\langle k,l\vert,
\end{align}
it is easy to show that the proper (anti-)symmetrization of the Hilbert
matrix elements 
\begin{align}
\alpha_{i,j}(t)=\begin{cases}
\dfrac{\beta_{i,j}(t)+g\cdot\beta_{j,i}(t)}{\sqrt{2}}=\sqrt{2}\beta_{i,j}(t) & \forall\, i<j\\[1em]
\beta_{i,i}(t)\dfrac{(1+g)}{2} & \forall\, i=j
\end{cases}
\end{align}
gives exactly the expected results in the Fock space (see Eqs. (\ref{eq:Dens_bos})
and (\ref{eq:Dens_fer})). 

\subsection{Occupation numbers}

The expectation value of the number operator $\hat{n}_{k}$, corresponding
to the average number of particles in the mode $k$, can be calculated
as follows: 
\begin{align}
\left\langle n_{k}\right\rangle  & =\mathrm{Tr}[\rho^{\mathscr{F}}(t)\hat{n}_{k}]=\mathrm{Tr}[\rho^{\mathscr{F}}(t)\hat{c}_{k}^{\dagger}\hat{c}_{k}]\nonumber\\
 & =\sum_{i}{}_{g}\langle i,i\vert \rho^{\mathscr{F}}(t)\hat{c}_{k}^{\dagger}\hat{c}_{k}\vert i,i \rangle _{g}+\sum_{i,j>i}{}_{g} \langle i,j\vert \rho^{\mathscr{F}}(t)\hat{c}_{k}^{\dagger}\hat{c}_{k}\vert i,j \rangle _{g}\nonumber\\
 & =\sqrt{2}\sqrt{2}\rho_{k,k;k,k}^{\mathscr{F}}(t)+\sum_{i<k}{}_{g}\langle i,k \vert \rho^{\mathcal{\mathscr{F}}}(t)\hat{c}_{k}^{\dagger}\hat{c}_{k}\vert i,k \rangle _{g}+\sum_{j>k}{}_{g}\langle k,j\vert\rho^{\mathcal{\mathscr{F}}}(t)\hat{c}_{k}^{\dagger}\hat{c}_{k}\vert k,j \rangle _{g}\nonumber\\
 & =2\rho_{k,k;k,k}^{\mathscr{F}}(t)+\sum_{i<k}\rho_{i,k;i,k}^{\mathscr{F}}(t)+\sum_{j>k}\rho_{k,j;k,j}^{\mathscr{F}}(t).
\end{align}
Upon recalling that in the Hilbert space the symmetry exchange requires
\begin{align}
\rho_{i,j;k,l}^{\mathscr{H}}(t)=g\cdot\rho_{j,i;k,l}^{\mathcal{\mathscr{H}}}(t)=g\cdot\rho_{i,j;l,k}^{\mathcal{\mathscr{H}}}(t),
\end{align}
we may rewrite $\left\langle n_{k}\right\rangle $ in the Hilbert
space, also using Eq. (\ref{eq:Dens_fer}), as follows: 
\begin{align}
\left\langle n_{k}\right\rangle  & =2\rho_{k,k;k,k}^{\mathcal{\mathscr{H}}}(t)+\sum_{i<k}2\rho_{i,k;i,k}^{\mathscr{H}}(t)+\sum_{j>k}2\rho_{k,j;k,j}^{\mathscr{H}}(t)\nonumber\\
 & =2\left(\rho_{k,k;k,k}^{\mathscr{H}}(t)+\sum_{i\neq k}\rho_{i,k;i,k}^{\mathcal{\mathscr{H}}}(t)\right)=2\sum_{i}\rho_{i,k;i,k}^{\mathscr{H}}(t),
\end{align}
since $\rho_{k,j;k,j}^{\mathscr{H}}=g^{2}\rho_{j,k;j,k}^{\mathscr{H}}=\rho_{j,k;j,k}^{\mathscr{H}}$.

\subsection{Entropies}
Given a quantum system, it is natural to ask how to measure the amount of quantum correlations in it. Besides representing an intriguing trait of quantum mechanics, quantum entanglement has turned into a fundamental resource for quantum information theory and quantum computing, since it can be used to implement protocols and tasks that could not be accomplished within the classical framework \cite{hor09}. The term entanglement refers to an intrinsic relation between subsystems of a composite quantum system: in an entangled state, each subsystem cannot be described independently of the state of the other one, or, in other words, what we know (ignore) about A, is what we know (ignore) about B, and vice-versa.\\
For a system composed of two subsystems A and B (bipartite) described by the density matrix $\rho_{AB}$, the entanglement among A and B can be quantified in different ways \cite{gun09}, depending on the reduced state of the subsystem $\rho_{A(B)}$ and the size $d_{A(B)}$ of the subsystem $A(B)$. In the case of pure states, the entanglement can always be measured with the von Neumann entropy $\mathcal{S}(\rho_{A})=-\rho_A$log$_2\rho_A$, with $\mathcal{S}(\rho_{A}) = \mathcal{S}(\rho_{B})$, \cite{nielsenchuang}. \\
Here we consider a compound system described the total density matrix $\rho_{SE}$, where a quantum system S is coupled to an external bath E, acting as a noise source. In this case the entanglement between system and environment
gives a measure of the decoherence, which quantifies the loss of coherence
in the quantum correlations of the system \cite{zur03}. In this picture, decoherence
can be evaluated via the von Neumann entropy of the quantum system $\mathcal{S}(\rho_{S})$, with $\rho_{S}=\textrm{Tr}_E\: \rho$.
Whenever the quantum system S contains indistinguishable particles, this quantity should be evaluated in the Fock space, which is the
natural space for the system since it accounts for the exchange symmetry.
Indeed, we have 
\begin{align}
\mathcal{S}(\rho^{\mathscr{F}}_S)=-\frac{1}{\ln(d_{g})}\mathrm{Tr}[\rho^{\mathscr{F}}_S\ln\rho^{\mathcal{\mathscr{F}}}_S]>-\frac{1}{\ln(K^2)}\mathrm{Tr}[\rho^{\mathscr{H}}_S\ln\rho^{\mathcal{\mathscr{H}}}_S]=\mathcal{S}(\rho^{\mathcal{\mathscr{H}}}_S),
\end{align}
since $\rho^{\mathscr{H}}_S$ and $\rho^{\mathscr{F}}_S$ have the same
eigenvalues: they only differ for a unitary transformation, and the
additional eigenvalues of $\rho^{\mathcal{\mathscr{H}}}_S$ are zeros
that do not contribute to the entropy. We therefore conclude that
$\mathrm{Tr}[\rho^{\mathscr{H}}_S\ln\rho^{\mathcal{\mathscr{H}}}_S]=\mathrm{Tr}[\rho^{\mathscr{F}}_S\ln\rho^{\mathcal{\mathscr{F}}}_S]$,
and the only difference between $\mathcal{S}(\rho^{\mathcal{\mathscr{F}}}_S)$
and $\mathcal{S}(\rho^{\mathscr{H}}_S)$ is given by different normalization
($d_{g}=\frac{K(K+g)}{2}<K^2$): so we conclude that $\mathcal{S}(\rho^{\mathscr{H}}_S)$
underestimates the loss of quantum correlations with respect to $\mathcal{S}(\rho^{\mathscr{F}}_S)$.
The reason is intuitively obvious: since the system always possesses
a residual amount of correlations due to exchange symmetry, these
correlations are seen as quantum correlations by the entropy of the
Hilbert space, which is devised for distinguishable particles. On
the other hand, they are correctly not counted by the von Neumann
entropy evaluated in the Fock space. Indeed, they are not genuine
quantum correlations - like entanglement or quantum discord \cite{oll01}- which may be
exploited to perform quantum information tasks. 

\section{Guidelines for Numerical implementation}

\label{s:num}

\subsection{Base ordering and indexing}

One of the main problems in implementing numerically the calculations
of operators is the different indexing in Hilbert and Fock spaces.
This situation is made more involved by the differences between allowed
states for fermions and bosons. Let us see this with an example. Let
us consider a system with $N=2$ identical particles, which can occupy
$K=4$ sites, or modes. The allowed states in the Hilbert and Fock
spaces are given in Table \ref{tab:HF_basis_set}. 
\begin{table}[h!]
\begin{centering}\renewcommand\arraystretch{1.25}
\begin{tabular}{cccc}
\toprule
\textbf{Space} & \textbf{Basis set} & \textbf{Dimension} \tabularnewline \hline\hline
$\begin{array}{c}
\displaystyle{\mathscr{H}_{2}}\\
\textit{\small(distinguishable)}
\end{array}$
& $\begin{array}{llll}
\left|1,1\right\rangle _{\phantom{s}}& \left|1,2\right\rangle _{\phantom{s}} &\left|1,3\right\rangle _{\phantom{s}} &\left|1,4\right\rangle _{\phantom{s}}\\
\left|2,1\right\rangle & \left|2,2\right\rangle  &\left|2,3\right\rangle  &\left|2,4\right\rangle \\
\left|3,1\right\rangle  &\left|3,2\right\rangle  &\left|3,3\right\rangle  &\left|3,4\right\rangle \\
\left|4,1\right\rangle  &\left|4,2\right\rangle  &\left|4,3\right\rangle  &\left|4,4\right\rangle 
\end{array}$ & 16  \tabularnewline \hline
$\begin{array}{c}
\displaystyle{\mathscr{F}_{2}^{B}}\\
\textit{\small(bosons)}
\end{array}$
 & $\begin{array}{llll}
\left|1,1\right\rangle _{s} &\left|1,2\right\rangle _{s} &\left|1,3\right\rangle _{s} &\left|1,4\right\rangle _{s}\\
& \left|2,2\right\rangle _{s} &\left|2,3\right\rangle _{s} & \left|2,4\right\rangle _{s}\\
&&\left|3,3\right\rangle _{s} &\left|3,4\right\rangle _{s}\\
&&& \left|4,4\right\rangle _{s} 
\end{array}$ & 10\tabularnewline \hline
$\begin{array}{c}
\displaystyle{\mathscr{F}_{2}^{F}} \\
\textit{\small(fermions)} 
\end{array}$
& $\begin{array}{llll}
\phantom{\left|1,1\right\rangle _{a}} &\left|1,2\right\rangle _{a} &\left|1,3\right\rangle _{a} &\left|1,4\right\rangle _{a}\\
&&\left|2,3\right\rangle _{a} & \left|2,4\right\rangle _{a}\\
&&&\left|3,4\right\rangle _{a} \\
\end{array}$& 6\tabularnewline \bottomrule 
\end{tabular}
\par\end{centering}

\caption{Basis sets for Hilbert and Fock spaces of $N=2$ identical particles,
which can occupy $K=4$ sites.}
\label{tab:HF_basis_set}
\end{table}

Since any vector or matrix must be indexed with a progressive index
$m$, we have to define a global index $m$ that depends on the single-particle
states $i$ and $j$ and follows the correct ordering when
basis set states $\left|m\right\rangle $ are $\left|i,j\right\rangle $,
$\left|i,j\right\rangle _{s}$ or $\left|i,j\right\rangle _{a}$.
It turns out that we have
\begin{align}
\mathscr{H}_{2}: & \quad m=K(i-1)+j,\label{mijd}\\
\mathscr{F}_{2}^{B/F}: & \quad m=K(i-1)+j-s(g,i)\,,\label{mijbf}
\end{align}
where $s(g,i)$ is a correction term that takes into account the fact
that states with indices exchanged must not be counted again in Fock space,
and also that states with identical indices are forbidden for fermions
($g=-1$). From an intuitive point of view, we can think that $i$
and $j$ in $\left|i,j\right\rangle $ are two numbers living on a
ring $\mathbb{Z}_{K}=\{1,2,...,K\}$: $i$ plays the role of the tens,
while $j$ plays the role of units and, overall, we have $m=K(i-1)+j$.
By a simple combinatorial reasoning we find: 
\begin{align}
s(g,i)=\frac{i(i-g)}{2}.
\label{eq:corr_term_s}
\end{align}
Indeed, for a fixed value of $i$, denoted as $i^\ast$, the number of forbidden states which must
be subtracted from $m$ is
\begin{align}
B:\quad &\operatorname{card}\{ \vert i,j\rangle \mid i\leq i^\ast \land j<i \}=\sum_{i=1}^{i^\ast} (i-1)=\sum_{i=0}^{i^\ast-1} i \, ,\\
F:\quad &\operatorname{card}\{ \vert i,j\rangle \mid i\leq i^\ast \land j\leq i \}=\sum_{i=1}^{i^\ast} i =\sum_{i=0}^{i^\ast} i \, .
\end{align}
In both cases, we calculate the result with the
Gauss formula $\sum_{i=0}^{n}i=\frac{1}{2}n(n+1)$.
One can easily verify that $\{\left|i,j\right\rangle _{s(a)} \mid i\leq i^\ast \land j\leoq i \}$ are exactly the states not appearing in Table \ref{tab:HF_basis_set} since forbidden.

So, according to Eqs. (\ref{mijbf}) and (\ref{eq:corr_term_s}), the state $\left|3,4\right\rangle$, e.g., is the basis state $\left|m\right\rangle =\left|12\right\rangle $
in $\mathscr{H}_{2}$, the basis state $\left|m\right\rangle =\left|9\right\rangle $
in $\mathscr{F}_{2}^{B}$, and the basis state $\left|m\right\rangle =\left|6\right\rangle $
in $\mathscr{F}_{2}^{F}$. This allows us to scan all the elements
of vector states and operators in terms of the single-particle states
$i$ and $j$, and it also lets us to switch easily from their Hilbert
representation to the Fock one and vice versa. 

\subsection{Reshaping cycle}

\label{ss:res} Now, it is worth noting that, in order to fill-in
properly the elements of an operator $O$ in the space $\mathscr{F}_{2}^{B/F}$,
starting from the corresponding operator in the Hilbert space (the
so-called \emph{reshaping} operation), we must use cycles like \begin{lyxcode}
for~i=1,N \begin{lyxcode} for~j=i+$\Delta$,N \begin{lyxcode}
for~k=1,N \begin{lyxcode} for~l=k+$\Delta$,N \begin{lyxcode}
$O^{\mathscr{F}}(i,j;k,l)=O^{\mathscr{H}}(i,j;k,l)\cdot...$ \end{lyxcode}
end \end{lyxcode} end \end{lyxcode} end \end{lyxcode} end \end{lyxcode}
where the correction 
\begin{align}
\Delta=\frac{1-g}{2}
\label{eq:delta}
\end{align}
is 0 for bosons (i.e., states with $i=j$ are allowed) and 1 for fermions
(i.e., states with $i=j$ are forbidden). 

\subsection{Computational and storage considerations}

Working in the Hilbert space offers an obvious advantage from the physical
point of view, since one has a clear identification of the degrees of
freedom associated to each particle, and a better indexing of states.
On the other hand, a couple of issues arises from the point of view
of numerical implementation. The first is linked to the larger dimension
of the space and may be properly addressed by opportunely inverting Eqs.
(\ref{mijbf}) and using reshaping cycles as those presented in Section
\ref{ss:res}, so that the number of operations is not significantly
larger in the Hilbert space \cite{NumbOper}. Let us define the function
\begin{align}
f_{\hbox{\tiny K}}^{\hbox{\tiny g}}(r)&:=
\begin{cases}
\displaystyle{m-1-\sum_{n=K-\Delta-r+1}^{K-\Delta} n} & \text{for } r \in \mathbb{N}^+\\
m-1 & \text{for } r=0
\end{cases}\nonumber\\
&=m-1-\frac{r}{2}\left(2K+g-r \right) \,\,\,\,\,\,\text{ for } r \in \mathbb{N}\, ,
\label{eq:fgK}
\end{align}
where $\Delta$ is defined in Eq. (\ref{eq:delta}), $g=\,\pmo1$ for bosons (fermions), and $K$ is the number of modes of the quantum system. The expression in Eq. (\ref{eq:fgK}), which is in principle the result for $r\in\mathbb{N}^+$, returns $m-1$ for $r=0$, thus it already summarizes the two distinct cases. Let $\bar{r}_{\hbox{\tiny K}}^{\hbox{\tiny g}}$ be the greater value of $r\in\{0,1,...,K-1\}$ such that $f_{\hbox{\tiny K}}^{\hbox{\tiny g}}(r)\geq 0$, i.e.
\begin{equation}
\bar{r}_{\hbox{\tiny K}}^{\hbox{\tiny g}} = \max \{r\mid r\in\{0,1,...,K-1\} \land f_{\hbox{\tiny K}}^{\hbox{\tiny g}}(r)\geq 0 \}.
\end{equation}
Hence, the inverse formulae of Eqs. (\ref{mijbf}) are given by 
\begin{equation}
\renewcommand\arraystretch{1.5}
\left\lbrace
\displaystyle{
\begin{array}{lll}
i_{\hbox{\tiny K}}^{\hbox{\tiny g}}(m) &=&1+\bar{r}_{\hbox{\tiny K}}^{\hbox{\tiny g}}\\
j_{\hbox{\tiny K}}^{\hbox{\tiny g}}(m) &=& \Delta+i_{\hbox{\tiny K}}^{\hbox{\tiny g}}(m)+f_{\hbox{\tiny K}}^{g}(\bar{r}_{\hbox{\tiny K}}^{\hbox{\tiny g}}) \, .
\end{array}}\right.
\end{equation}
The other and major issue is instead related to
the storage of the matrix elements of states and operators, since
in both cases using the Hilbert space description amounts to storing
several empty cells, i.e. those corresponding to states with the wrong
symmetry. This problem may be addressed by exploiting the above mapping,
and also noticing that the involved matrices are often sparse, e.g.
when systems with only nearest-neighbour interactions are considered
\cite{cpc17}, so that sparse matrix declarations and algorithms
may be exploited to reduce the storage space. 

\section{Concluding remarks}

\label{s:out} In graduate physics courses, second quantization and
the Fock space are presented as the natural framework to deal with
quantum systems made of many indistinguishable particles, leaving
the impression that the Hilbert space description may be left behind.
While this is certainly true for the description of quantum states
of those systems, the evaluation of some specific observable or the
study of the system dynamics may be often more conveniently pursued
using the Hilbert space description. 

A research-oriented teaching of these topics should therefore reflect
the importance of both descriptions, and provide tools to connect
them in the most straightforward way. To this aim, we have provided
here a gentle and self-contained introduction to details of the transformation
rules between the different description of states and operators in
the two spaces. In particular, we have devoted some attention to the
two-particle case, since this already contains most of the interesting
features related to indistinguishability. The paper aims at being
a concise reference about the different representations for students
and researchers working with systems made of many identical particles,
especially those interested in numerical approaches to the system
dynamics. 

\section*{Acknowledegements}

This work has been supported by JSPS through FY2017 program (grant
S17118) and by SERB through the VAJRA award (grant VJR/2017/000011).
PB and MGAP are members of GNFM-INdAM. 



\begin{thebibliography}{10}
\bibitem{APE} It may be useful to remind here an aphorism by Asher
Peres:\textquote{Quantum phenomena do not occur in a Hilbert space.
They occur in a laboratory}.
\bibitem{Decomp} This decomposition holds only for $N=2$, since
for higher dimensions there is an additional subspace $\mathscr{N}$
containing states that do not possess any symmetry, but are required
to produce a complete basis set for $\mathscr{H}_{N}$.
\bibitem{Alt_dyn} Considering that $[\hat{H},\hat{S}]=0$, we can
obtain the same results with an alternative approach: we begin with
a non-symmetrized state $\left|\Psi\right\rangle $, evolve it with
the distinguishable-particle Hamiltonian $\hat{H}$, and then perform
the symmetrization at the end of the evolution, since $\hat{H}_{s}\left|\Psi\right\rangle _{s}=\hat{H}\left|\Psi\right\rangle _{s}=\hat{H}\hat{S}\left|\Psi\right\rangle =\hat{S}\hat{H}\left|\Psi\right\rangle $.
\bibitem{NumbOper} The issue concerning the larger number of operations cannot be avoided when using built-in subroutines, such as matrix multiplication and diagonalization.
\end{thebibliography}
\end{document}